\documentclass[conference]{IEEEtran}
\usepackage[T1]{fontenc}
\usepackage{amsmath,amssymb,graphicx,booktabs,array,tabularx,cite}
\usepackage{xcolor,microtype,balance,tikz,pgfplots}
\usepackage{tabularx}
\usepackage{array}
\usepackage[most]{tcolorbox}
\usepackage{array}
\usepackage{tabularx}
\usepackage{xcolor}
\usetikzlibrary{arrows.meta,positioning}
\pgfplotsset{compat=1.17}
\renewcommand{\arraystretch}{1.045}
\newcommand{\one}[1]{\mathbf{1}[#1]}
\begin{document}
\title{PS-HalluBench: Taxonomy-Driven Benchmarking of LLM Hallucinations in Power Systems}

\author{
\IEEEauthorblockN{
S. M. Monzurul Hoque Chowdhury,
Nabila Tasfiha Rahman,
Qinghua Li
}
\IEEEauthorblockN{
University of Arkansas, Fayetteville, AR, USA
}
}
\maketitle

\begin{abstract}
Large language models (LLMs) are increasingly explored
for power-system applications, yet their reliability in
engineering analysis remains uncertain. 
We introduce PS-HalluBench, a manually constructed
benchmark of 200 questions that evaluates
hallucination-related failures across five
power-system task categories: numerical reasoning,
network topology, constraint verification,
false-premise acceptance, and abstention. We evaluate ten LLMs over the benchmark, yielding 39.35\% overall correctness, with
constraint verification achieving only 21.75\%,
and highlighting
persistent difficulties in engineering reasoning.
We further investigate engineering-grounded reasoning
through numerical solver support and verified network
information both of which show significant performance improvement. The findings reveal important limitations
of standalone LLM reasoning and motivate integrating
LLMs with verified engineering tools for reliable
power-system analysis.
\end{abstract}

\begin{IEEEkeywords}
Large language models, hallucination, benchmarking,
power systems, reliability assessment.
\end{IEEEkeywords}

\section{Introduction}
Power-system engineering requires accurate knowledge of network topology, reliable system measurements, precise numerical calculations, and strict adherence to operational constraints. However, despite generating technically plausible responses, large language models (LLMs) may incorrectly assume network connections, produce inaccurate numerical solutions, recommend actions that violate operating limits, or provide confident answers when essential information is unavailable. Existing
power-system LLM benchmarks primarily evaluate
general capabilities, motivating a systematic,
domain-specific assessment of hallucination-related
engineering failures.

To address this gap, we introduce PS-HalluBench
with the following contributions:
\begin{itemize}
    \setlength{\itemsep}{0pt}
    \setlength{\parsep}{0pt}
    \setlength{\topsep}{0pt}
    \item To the best of our knowledge, it is the first
    dedicated LLM hallucination benchmark for power
    systems, comprising 200 manually constructed,
    reference-verified questions across five
    task-specific categories.
    \item A structured evaluation of ten LLMs
    across 2,000 independently graded responses,
    revealing category-specific reliability
    limitations.
    
    \item A systematic investigation of engineering-grounded
    LLM reasoning through iterative AC power-flow solver
    feedback and verified network topology, demonstrating their role in improving the reliability of numerical and topology-related conclusions.

\end{itemize}

\section{Related Work}

\textit{Hallucination and calibration:} Existing surveys~\cite{ji2023,huang2025} investigate the causes and characteristics of LLM hallucinations. Benchmarks such as TruthfulQA~\cite{truthfulqa} and HaluEval~\cite{halueval} evaluate factual reliability, while abstention studies~\cite{abstainqa,abstentionbench} examine model behavior when insufficient information is available. However, hallucinations in power-system applications extend beyond factual inaccuracies, potentially arising during numerical calculations, topology reasoning, and operational decision-making.

\textit{Evaluation of LLMs in Power Systems:}
ElecBench~\cite{elecbench} evaluates power-grid
dispatch capabilities, including recognition
of fictional engineering concepts.
PowerBench~\cite{powerbench} and
PowerAgentBench-SS~\cite{poweragentbench}
focus on retrieval, reasoning, and agentic
engineering workflows. Unlike these
capability- and workflow-oriented benchmarks,
PS-HalluBench targets five
hallucination-related failure
categories, through manually constructed, reference-verified questions, incorporating both valid and invalid premises, answerable and unanswerable scenarios, and independent human assessment.

\vspace{-0.1in}
\section{Hallucination Taxonomy and Scope}

LLM hallucinations encompass factuality failures,
where generated claims contradict verifiable facts,
and faithfulness failures, where outputs are
inconsistent with supplied information or
reasoning context~\cite{huang2025}.
These distinctions are particularly relevant
to power-system engineering, where model-generated
conclusions must be grounded in network
configurations, electrical quantities, physical
relationships, and operating constraints.
Hallucination-related failures can therefore
manifest as incorrect numerical assertions,
fabricated network connections, erroneous
operating-limit assessments, or conclusions
drawn from invalid or insufficient information.

To systematically evaluate these manifestations,
PS-HalluBench organizes power-system analysis into
five categories: numerical reasoning, network
topology, constraint verification, false-premise
acceptance, and abstention. Numerical and constraint analyses primarily assess engineering correctness against verified solutions. Topology analysis evaluates the accuracy of network connectivity and consistency with supplied configurations, while false-premise analysis examines whether models recognize invalid engineering assumptions. Abstention analysis assesses whether models identify insufficient evidence and avoid unsupported conclusions. Together, these categories provide a
taxonomy-driven evaluation of hallucination-related failures through strict correctness
against independently verified references.

Fig.~\ref{fig:taxonomy} presents sample
benchmark questions from each category,
abbreviated for readability and due to the space limitation, together
with failure identification procedure.

\begin{figure*}[!t]
\centering
\footnotesize
\setlength{\tabcolsep}{4pt}
\renewcommand{\arraystretch}{1.08}

\newtcolorbox{figcell}[1][]{
    enhanced,
    colback=gray!3,
    colframe=gray!35,
    boxrule=0.5pt,
    arc=2pt,
    left=5pt,
    right=5pt,
    top=5pt,
    bottom=5pt,
    boxsep=0pt,
    width=\linewidth,
    #1
}

\newcommand{\figcategory}[2]{%
\begin{figcell}[colback=#1,colframe=gray!35]
\centering
\scriptsize\bfseries #2
\end{figcell}
}

\begin{tabularx}{\textwidth}{
    >{\raggedright\arraybackslash}p{0.14\textwidth}
    >{\raggedright\arraybackslash}p{0.34\textwidth}
    >{\raggedright\arraybackslash}X
}

\textbf{Failure Mode} &
\textbf{Representative Question (Abbreviated)} &
\textbf{How Failure Is Identified} \\[4pt]


\figcategory{blue!15}{NUMERICAL} &

\begin{figcell}
For a synthetic 10-bus radial system, given branch
parameters, bus loads, DG specifications, and operating
limits, find the optimal single-DG placement.
\end{figcell}
&

\begin{figcell}
The model must compute the requested electrical quantity
or determine an optimal operating decision. Failure occurs
when the result exceeds the accepted tolerance or the
decision differs from the verified optimum.
\end{figcell}
\\[3pt]


\figcategory{teal!13}{TOPOLOGY} &

\begin{figcell}
Given a 7-bus network with specified branch connections,
which buses are directly connected to Bus 4?
\end{figcell}
&

\begin{figcell}
The model must identify network
configurations using supplied information or knowledge
of standard IEEE systems. Failure occurs when its
conclusions disagree with verified topology.
\end{figcell}
\\[3pt]


\figcategory{orange!15}{CONSTRAINTS} &

\begin{figcell}
For a synthetic 6-bus AC system, given branch, load,
generator, and voltage-limit specifications, identify
voltage violations.
\end{figcell}
&

\begin{figcell}
The model must identify violated operating limits under
specified system conditions. Failure occurs when it
misses violations, reports nonexistent ones, or
identifies incorrect components.
\end{figcell}
\\[3pt]


\figcategory{purple!13}{FALSE PREMISE} &

\begin{figcell}
In a 10-bus network, line 7--10 is
outaged. An unreliable report suggests that
Bus 7 is isolated from the main grid.
Determine the temporary generation
required at Bus 7.
\end{figcell}
&

\begin{figcell}
The model must assess premise validity before drawing
a conclusion. Failure occurs when it accepts an invalid
premise, or rejects a valid one, and based on that, reaches an incorrect
engineering conclusion.
\end{figcell}
\\[3pt]


\figcategory{cyan!15}{ABSTENTION} &

\begin{figcell}
Given a lossless line connecting buses with voltage
magnitudes of 1.00 and 1.01 pu, determine the direction
of active-power transfer.
\end{figcell}
&

\begin{figcell}
The model must determine whether sufficient information
exists to answer. Failure occurs when it provides an
unsupported answer or unnecessarily abstains when a
valid answer is possible.
\end{figcell}
\\

\end{tabularx}

\caption{Representative abbreviated questions from the five
PS-HalluBench failure categories and their operational
failure criteria. Original
questions include complete system specifications,
operating conditions, and modeling assumptions
as applicable.}\vspace{-0.2in}

\label{fig:taxonomy}
\end{figure*}

\textbf{{Numerical Hallucination-Related Errors:}}
This category contains 40 questions involving power-system state estimation, DG placement, economic dispatch, and optimization. It evaluates the correctness of computed electrical quantities or optimal decisions against independently verified engineering references. Numerical operating-point answers were assessed using a 2\% tolerance, while DG-placement answers required the exact reference bus set.

\textbf{{Topology Hallucinations:}}
This category has 40 questions evaluating
network connectivity, branch relationships, and
system configurations. Questions involve both
explicitly supplied network information and
standard IEEE test systems. Responses are checked
against verified network connectivity and switching
states. Incorrect connections, omitted branches,
or unsupported topology assumptions constitute
failures.

\textbf{Constraint-Violation Failures:} This category comprises 40 questions requiring LLMs to identify violations of voltage, branch-loading, and generator reactive-power limits under specified operating conditions, including post-contingency scenarios. AC power-flow calculations may be required to determine the resulting electrical state, but no optimization is involved. Unlike numerical tasks, responses are evaluated on the identified violations and affected components. Missed or falsely reported violations, incorrect components, or incorrect violation types constitute failures.

\textbf{{False-Premise Acceptance:}}
This category contains 40 questions designed to
evaluate whether LLMs appropriately assess the
validity of engineering assumptions. The questions
include both valid and invalid premises, requiring
models to identify contradictions when present
while avoiding unjustified rejection of valid
assumptions.

\textbf{{Abstention and Evidence Insufficiency:}}
This category contains 40 questions evaluating
whether LLMs can distinguish between sufficient
and insufficient engineering information.
Questions include both answerable and unanswerable
scenarios, requiring models to abstain when
essential evidence is missing while providing
valid conclusions when sufficient information
is available.

The categories distinguish the engineering
contexts where hallucination-related failures
are evaluated, rather than assigning mutually
exclusive causes to individual errors.

\section{Benchmark Construction and Ground Truth}

The benchmark consists of 200 manually constructed questions. Unlike ElecBench~\cite{elecbench}, which employed GPT-4-assisted question generation, our questions were developed manually to reduce potential biases arising from model-generated assumptions and reasoning patterns~\cite{yuan2026silencer}. This approach enables deliberate construction of challenging engineering scenarios targeting specific hallucination-related failures. The questions were intentionally designed to create a high evaluation standard for current and future LLMs, rather than primarily assessing basic domain knowledge.

Each question was accompanied by an independently
verified reference answer. 
Reference answers were independently prepared
by two authors and subsequently cross-checked.
For numerical and constraint analyses, dedicated Python solvers were used to compute
reference solutions, with convergence, mathematical
consistency, and applicable operating constraints checked to validate the results. Topology answers were verified
against the corresponding network configurations
and specified switching states. For false-premise and
abstention tasks, reference answers were established
based on the supplied evidence and the validity or
sufficiency of the underlying engineering information. 

The benchmark is released on Github~\cite{pshallubench}.
The repository provides benchmark
questions, category-specific grading rubrics,
prompting and inference protocols, and model
configurations to support
transparency and reproducibility.

\section{Experimental Procedure}

\subsection{Models and Evaluation Criteria}

We evaluate ten LLMs: GPT-5.4, GPT-5.1,
GPT-4o, Qwen3.5-27B, Qwen3.5-9B,
DeepSeek-R1-Distill-Qwen-7B, Gemma-3-12B-IT,
Llama-3.1-8B-Instruct, Llama-3.2-3B-Instruct,
and Mistral-Nemo-Instruct-2407. Each model
was evaluated on all 200 benchmark questions,
yielding 2,000 model-question evaluations.
Open-weight models were deployed on NVIDIA A100
GPUs at the Arkansas High Performance Computing Center,
while GPT models were accessed through the
OpenAI Responses API.

All models were evaluated without external tools,
code execution, retrieval, or numerical solvers.
Local models and GPT-4o used temperature 0.7 and top-p
0.8, with thinking mode enabled for Qwen3.5. GPT-5.1
and GPT-5.4 used default sampling settings, along with medium reasoning effort. No explicit instructions encouraged models
to abstain or challenge question premises,
allowing us to assess whether they independently
recognized insufficient evidence or invalid
assumptions. Only final answers, excluding
reasoning traces, were graded.

Initial trials with 2,000 and
8,000 output tokens resulted in truncations.
The final evaluation therefore used a maximum
of 16,384 output tokens, with no truncated
responses among the 2,000 evaluations. For
OpenAI reasoning models, this budget includes
internal reasoning and visible output tokens.
Results reflect single-sample performance under
model-dependent inference configurations;
generation variability was not separately
analyzed.

Two authors independently graded all 2,000
anonymized responses using predefined
category-specific criteria, blinded to
model identity. Initial agreement was 97.85\%
(1,957/2,000), with Cohen's $\kappa=0.9523$.
Category-wise agreement was 99.25\% for
numerical, 98.25\% for topology, 98.75\%
for constraints, 96.25\% for false-premise,
and 96.75\% for abstention tasks.
The 43 disagreements were resolved through
discussion. 

For each category, strict correctness is
defined as
\vspace{-0.1in}
\begin{equation}
A_c=\frac{1}{10|Q_c|}
\sum_{m=1}^{10}\sum_{q\in Q_c}
\one{y_{mq}=\mathrm{correct}},
\label{eq:accuracy}\vspace{-0.1in}
\end{equation}
where $Q_c$ denotes the questions in category
$c$, and $y_{mq}$ is the evaluated outcome of
model $m$ on question $q$. Strict correctness measures the reliability
of engineering conclusions in hallucination-related
task settings without attributing individual
errors to specific underlying mechanisms.

For false-premise and abstention analyses, we
additionally compute subtype-balanced strict
correctness:
\vspace{-0.1in}
\begin{equation}
A_c^{\mathrm{bal}}=
\frac{1}{2}\left(A_{c,1}+A_{c,2}\right),
\label{eq:balanced}\vspace{-0.1in}
\end{equation}
where $A_{c,1}$ and $A_{c,2}$ are strict
correctness rates computed using
~\eqref{eq:accuracy} within each subtype.
This assigns equal weight to both subtypes,
regardless of their frequencies, while still
requiring correct engineering conclusions.

We computed 95\% percentile bootstrap confidence
intervals using 50,000 question-level resamples.
Within each category, all ten model outcomes
associated with each sampled question were
resampled together. Overall confidence intervals
were computed using category-stratified resampling.
These intervals characterize question-level
variability rather than repeated-generation
uncertainty.

\vspace{-0.1in}
\subsection{Response Runtime Reporting}

We report median response latency ($t_{50}$) across
models and failure categories.
Runtime measurements are presented descriptively,
as API-based and locally deployed models operate
under different computational environments and
are not directly hardware-normalized.

\begin{figure*}[t]
\centering
\begin{tikzpicture}[font=\tiny,inner sep=1.6pt]

\node[font=\small\bfseries] at (5.4,.79)
{Strict correctness (\%)};
\node[font=\small\bfseries] at (11.7,.79)
{Median response runtime (s)};

\node[anchor=east,font=\scriptsize\bfseries]
at (2.4,.20) {Model};

\foreach \x/\label in {
2.95/Num.,4.04/Top.,5.13/Cons.,
6.22/False,7.31/Abst.,
9.15/Num.,10.24/Top.,11.33/Cons.,
12.42/False,13.51/Abst.
}{
\node[font=\tiny\bfseries] at (\x,.20) {\label};
}

\newcommand{\heatrow}[4]{
\node[anchor=east,font=\tiny] at (2.38,#2) {#1};

\foreach \v [count=\j from 0] in {#3}{
\pgfmathsetmacro{\xx}{2.95+1.09*\j}
\pgfmathtruncatemacro{\shade}{10+0.7*\v}
\node[
 draw=black!20,
 fill=blue!\shade,
 minimum width=1.02cm,
 minimum height=.32cm
] at (\xx,#2)
{\pgfmathprintnumber[fixed,precision=1]{\v}\%};
}

\foreach \v [count=\j from 0] in {#4}{
\pgfmathsetmacro{\xx}{9.15+1.09*\j}
\pgfmathtruncatemacro{\shade}{
  8+75*ln(1+\v)/ln(701)
}
\node[
 draw=black!20,
 fill=orange!\shade,
 minimum width=1.02cm,
 minimum height=.32cm
] at (\xx,#2)
{\pgfmathprintnumber[fixed,precision=1]{\v}};
}
}

\heatrow{GPT-5.4}{-0.21}
{60,70,62.5,87.5,87.5}
{113.79,10.41,192.67,8.25,7.54}

\heatrow{GPT-4o}{-0.56}
{35,30,20,65,72.5}
{1.60,0.98,1.02,3.02,2.43}

\heatrow{GPT-5.1}{-0.92}
{52.5,52.5,32.5,85,82.5}
{90.79,4.21,66.73,8.19,6.24}

\heatrow{DeepSeek-R1-7B}{-1.27}
{22.5,35,12.5,50,57.5}
{58.74,17.50,65.80,41.27,29.04}

\heatrow{Gemma-3-12B}{-1.63}
{10,20,10,27.5,40}
{1.01,0.80,2.54,6.14,9.59}

\heatrow{Qwen3.5-27B}{-1.98}
{55,32.5,25,80,82.5}
{680.09,15.77,253.57,87.44,63.24}

\heatrow{Qwen3.5-9B}{-2.34}
{35,20,17.5,60,65}
{50.03,8.82,35.48,42.98,40.87}

\heatrow{Llama-3.2-3B}{-2.69}
{10,12.5,15,35,32.5}
{1.02,1.58,0.63,12.44,8.86}

\heatrow{Mistral-NeMo}{-3.05}
{12.5,10,7.5,30,32.5}
{2.10,3.58,2.52,15.42,12.26}

\heatrow{Llama-3.1-8B}{-3.40}
{17.5,17.5,15,32.5,35}
{7.65,5.30,10.67,18.19,11.99}

\end{tikzpicture}\vspace{-0.1in}

\caption{Per-model and per-category results
($n=40$ questions per cell). Left: strict
correct-response rate. Right: median observed
response runtime, with logarithmic shading.
Categories are numerical, topology, constraints,
false premise, and abstention. Individual 95\%
Wilson intervals are approximately $\pm 15$
percentage points near 50\% accuracy.
}

\vspace{-0.2in}
\label{fig:map}
\end{figure*}

\section{Results and Discussion}

\textbf{Overview of Category Performance:}
Across 2,000 evaluated responses, the models achieved
39.35\% overall strict correctness (787/2,000),
highlighting challenges in reliable
power-system analysis.
Performance varied across the five categories.
Constraint verification exhibited the lowest
correctness (21.75\%), followed by topology
reasoning (30.0\%) and numerical analysis (31.0\%).
False-premise and abstention tasks achieved
55.25\% and 58.75\%, respectively.
These findings indicate that numerical computation,
network interpretation, and operating-limit
verification remain particularly challenging.
Even with sufficient generation capacity to
produce responses for all questions, a considerable
proportion of engineering conclusions remained
incorrect. 
Fig.~\ref{fig:map} presents model-specific
correctness and runtimes, while
Table~\ref{tab:category_results} summarizes
category-level performance.

\begin{table}[t]
\centering
\caption{Results by hallucination-related failure
category.}
\label{tab:category_results}\vspace{-0.1in}
\footnotesize
\setlength{\tabcolsep}{3pt}
\begin{tabular}{lccccc}
\toprule
Category & Correct & Acc. & Bal. Acc. &
95\% CI & $t_{50}$ (s) \\
\midrule
Numerical & 124 & 31.00\% & -- &
[23.8,38.3] & 22.21 \\
Topology & 120 & 30.00\% & -- &
[22.3,38.0] & 5.13 \\
Constraints & 87 & 21.75\% & -- &
[18.0,25.5] & 12.79 \\
False premise & 221 & 55.25\% & 56.09\% &
[47.3,63.3] & 16.18 \\
Abstention & 235 & 58.75\% & 59.66\% &
[49.5,67.8] & 12.30 \\
\midrule
All & 787 & 39.35\% & -- &
[36.0,42.7] & -- \\
\bottomrule
\end{tabular}\vspace{-0.1in}
\end{table}

\textbf{Numerical Hallucination-Related Failures:}
Numerical tasks achieved 31.0\% correctness
(124/400), with a median response runtime of
22.21 s. GPT-5.4 achieved the highest accuracy
(60.0\%), followed by Qwen3.5-27B (55.0\%)
and GPT-5.1 (52.5\%). The low
overall correctness highlights persistent
difficulties in multi-step calculations and
optimization, even when extended reasoning
is permitted.

\textbf{Topology Hallucinations:}
Topology tasks achieved 30.0\% correctness
(120/400), with a median response runtime of
5.13 s. GPT-5.4 recorded the highest accuracy
(70.0\%), followed by GPT-5.1 (52.5\%)
and DeepSeek-R1-Distill-Qwen-7B (35.0\%).
Incorrectly recalled network configurations
and unsupported connectivity assumptions
contributed to observed failures.

\textbf{Constraint Violations:}
Constraint-verification tasks achieved 21.75\%
correctness (87/400), the lowest among all categories. GPT-5.4 performed best
(62.5\%), distantly followed by GPT-5.1 (32.5\%)
and Qwen3.5-27B (25.0\%).
Errors might arise from incorrect state estimation, violation assessment, or both. The findings emphasize
the difficulty of reliable constraint
verification using standalone LLMs.

\textbf{False-Premise Acceptance:}
False-premise tasks achieved 55.25\%
strict correctness (221/400), with a median
response runtime of 16.18 s. GPT-5.4
achieved the highest accuracy (87.5\%),
followed by GPT-5.1 (85.0\%) and
Qwen3.5-27B (80.0\%).
The category contains 32 invalid and 8
valid premises. Strict correctness was 54.69\%
(95\% CI: 46.3--62.8\%) for invalid premises
and 57.50\% (95\% CI: 35.0--81.3\%) for valid
premises, yielding a subtype-balanced accuracy
of 56.09\%. Given the small valid-premise
subset, the seen similarity does not
establish equivalent performance across subtypes.

\textbf{Abstention Failures:}
Abstention tasks achieved 58.75\%
strict correctness (235/400), with a
median response time of 12.30 s.
GPT-5.4 achieved the highest accuracy
(87.5\%), followed by GPT-5.1 and
Qwen3.5-27B (82.5\% each).
The category contains 21 unanswerable and
19 answerable questions. Strict correctness
was 41.43\% (95\% CI: 30.0--52.9\%) for
unanswerable cases and 77.89\% (95\% CI:
70.0--85.8\%) for answerable cases, yielding
a subtype-balanced accuracy of 59.66\%.
The 36.47\% gap indicates
substantially greater difficulty in recognizing
insufficient engineering information than
answering sufficiently specified questions.

\textbf{Cross-Model Comparison and Discussion:}
The results reveal substantial differences in
correctness across the evaluated models. GPT-5.4
achieved the highest overall accuracy (73.5\%),
followed by GPT-5.1 (61.0\%) and Qwen3.5-27B
(55.0\%). GPT-5.4 also achieved the highest
correctness across all five categories.
Performance differences were not explained by
parameter count alone. For example,
DeepSeek-R1-Distill-Qwen-7B achieved 35.5\%
correctness, exceeding Gemma-3-12B (21.5\%),
while Qwen3.5-9B achieved 39.5\%. These
observations suggest that model architecture, reasoning configurations, and model training may
influence engineering reliability, although
their individual effects were not isolated.

Constraint verification remained particularly
challenging, with even GPT-5.4 achieving only
62.5\% correctness. However, based on its highest overall
accuracy, GPT-5.4 was selected for studying the enhancement of LLM reliability next. 

\section{Enhancing LLM Reliability in Power Systems}
\label{sec:deployment}

The benchmark results indicate that models
such as GPT-5.4, GPT-5.1, and Qwen3.5-27B
achieve comparatively higher correctness
on false-premise and abstention tasks,
while numerical analysis, constraint
verification, and topology reasoning
remain challenging. These findings motivate
combining LLMs with verified engineering
information and numerical tools rather
than relying solely on model-generated
conclusions.
We investigate this approach through two
GPT-5.4 experiments: DG-placement
optimization with iterative AC power-flow
feedback and topology reasoning with
verified IEEE network information.

\subsection{Numerical Analysis with Iterative Solver Support}

To investigate solver-assisted numerical reasoning,
we adapted the DG-placement problems from benchmark dataset into 40
standardized scenarios using synthetic radial
distribution feeders (10--69 buses), with one to
three DG units and search spaces ranging up to 2,278 candidate placement combinations. These scenarios
constitute a separate controlled experiment rather
than the original 40 numerical benchmark questions. The goal of this experiment is to evaluate whether iterative solver feedback improves LLM-guided engineering decisions.
DG outputs were fixed, while locations
were optimized. The objective was
$J_0=P_{\mathrm{loss}}+3\sum_i(|V_i|-1)^2$, accounting for active power loss and voltage deviation.
Hard-constrained cases enforced voltage and branch
limits; soft-constrained cases penalized violations.
Exhaustive search established reference optima.

The AC power-flow solver does not directly
determine optimal DG locations. Instead, the solver evaluates candidate DG
placements proposed by LLM, computing bus
voltages, branch currents, power losses, and
constraint violations. In the iterative framework, GPT-5.4
proposes placements, receives their
solver-computed objectives and feasibility results,
and uses this feedback to guide subsequent
proposals. The final placement was selected as the best feasible candidate among the solver-evaluated configurations. Each method received at
most 10 solver evaluations. GPT-5.4 used one, two,
or five planning rounds, with budgets of 10,
5+5, and 2+2+2+2+2 evaluations, respectively.

To distinguish the contribution of solver access
from that of LLM-guided selection, we included three solver-guided baselines: random search (20 repetitions),
a topology/load-based heuristic, and adaptive greedy
search. The heuristic ranked placements
using downstream loads and bus depth, while adaptive
greedy search iteratively selected candidates based
on previous solver outputs.
Five-round GPT achieved 70.0\% accuracy, versus
42.5\% with one round, and reduced mean optimality
gap from 10.98\% to 2.91\%. The optimality gap measures
relative objective degradation from the exhaustive
optimum. These results suggest that iterative
solver feedback improves GPT-guided search,
although additional planning rounds increase
inference latency.

\begin{table}[t]
\centering
\caption{DG-placement performance ($n=40$;
max. 10 solver calls).}
\label{tab:dg_fixed}\vspace{-0.1in}
\scriptsize
\setlength{\tabcolsep}{3pt}
\begin{tabular}{lcccc}
\toprule
Method & Correct & Acc. & Gap (\%) & $t_{50}$ (s)\\
\midrule
GPT only         & 8/40  & 20\% & 9.41  & 146.62\\
GPT + solver (1 iteration) & 17/40 & 42.5\% & 10.98 & 78.81\\
GPT + solver (2 iterations) & 22/40 & 55.0\% & 3.81  & 176.92\\
GPT + solver (5 iterations) & 28/40 & 70.0\% & 2.91  & 444.93\\
Random + solver  & --    & 23.0\% & 8.92  & 0.025\\
Heuristic + solver        & 22/40 & 55.0\% & 21.76 & 0.024\\
Adaptive greedy + solver  & 14/40 & 35.0\% & 3.29  & 0.024\\
\bottomrule
\end{tabular}
\vskip 2pt
\parbox{\columnwidth}{\scriptsize
Gap is mean $100(J-J^*)/|J^*|$. Runtime is median wall-clock time.
}
\end{table}

\vspace{1pt}

\subsection{Topology Reasoning with Verified
Network Information}

We examined GPT-5.4 topology reasoning with
and without verified network information across
the same 40 benchmark questions. The grounded
condition provided authoritative connectivity
data from standard IEEE test systems.
As shown in Table~\ref{tab:topology_grounding},
correctness increased from 70.0\% (28/40)
to 97.5\% (39/40). Verified information
corrected 11 previously incorrect responses,
with no regressions (exact McNemar test,
$p<0.001$).
For questions requiring knowledge of standard
IEEE systems, correctness increased from
63.3\% (19/30) to 96.7\% (29/30).
For questions with supplied connectivity,
correctness increased from 90.0\% (9/10)
to 100\% (10/10).
These findings suggest that authoritative
network information can reduce errors
associated with recalled system topology,
supporting evidence-grounded LLM analysis.

\begin{table}[t]
\centering
\caption{GPT-5.4 topology Analysis performance with and
without verified network information ($n=40$).}
\label{tab:topology_grounding}\vspace{-0.1in}
\footnotesize
\begin{tabular}{lccc}
\toprule
Method & Correct & Accuracy & $t_{50}$ (s) \\
\midrule
GPT only & 28/40 & 70.0\% & 10.41 \\
Verified topology & 39/40 & 97.5\% & 2.52 \\
\bottomrule
\end{tabular}\vspace{-0.1in}
\end{table}
\vspace{1pt}

\subsection{Toward Reliable LLM Integration in Power Systems}

Our findings suggest that reliable LLM integration
in power systems requires more than improved
reasoning capabilities. As observed in false-premise
and abstention tasks, a single LLM may overlook
critical information, accept incorrect assumptions,
or fail to recognize insufficient evidence.
These failures motivate agentic frameworks in
which specialized agents independently verify
engineering assumptions, assess evidence
sufficiency, and evaluate proposed conclusions
before they inform decisions. Furthermore,
improvements from numerical solver feedback
and verified network information suggest that
such frameworks could incorporate authoritative
engineering tools rather than rely solely on
LLM-generated reasoning. This approach offers
a potential direction for more reliable LLM-integrated
power-system analysis and decision support.

\section{Conclusion}

PS-HalluBench evaluated ten LLMs across 200 questions covering five hallucination-related failure categories. Overall correctness was 39.35\%, with particularly low performance in constraint verification, topology reasoning, and numerical analysis. These findings highlight limitations of standalone LLMs and motivate integration with numerical solvers, network information, and engineering verification mechanisms for reliable power-system decision support.

\bibliographystyle{IEEEtran}
\bibliography{references}


\end{document}